\documentclass[twocolumn,10pt]{article}
\usepackage{cite}
\usepackage{graphicx}
\usepackage{psfrag}
\usepackage{subfigure}
\usepackage{url}
\usepackage{amsmath}
\usepackage{algorithmic,algorithm}

\begin{document}

\author{Panos Papadimitratos and Aleksandar Jovanovic \\
            EPFL \\ 
            Switzerland \\
            Email: firstname.lastname@epfl.ch}

%
%


\title{GNSS-based Positioning: Attacks and Countermeasures}

\date{}
\maketitle


%



\begin{abstract}
Increasing numbers of mobile computing devices, user-portable, or embedded in vehicles, cargo containers, or the physical space, need to be aware of their location in order to provide a wide range of commercial services. Most often, mobile devices obtain their own location with the help of Global Navigation Satellite Systems (GNSS), integrating, for example, a Global Positioning System (GPS) receiver. Nonetheless, an adversary can compromise location-aware applications by attacking the GNSS-based positioning: It can forge navigation messages and mislead the receiver into calculating a fake location. In this paper, we analyze this vulnerability and propose and evaluate the effectiveness of countermeasures. First, we consider \emph{replay} attacks, which can be effective even in the presence of future cryptographic GNSS protection mechanisms. Then, we propose and analyze methods that allow GNSS receivers to detect the reception of signals generated by an adversary, and then reject fake locations calculated because of the attack. We consider three diverse defense mechanisms, all based on knowledge, in particular, own \emph{location}, \emph{time}, and \emph{Doppler shift}, receivers can obtain prior to the onset of an attack. We find that inertial mechanisms that estimate location can be defeated relatively easy. This is equally true for the mechanism that relies on clock readings from off-the-shelf devices; as a result, highly stable clocks could be needed. On the other hand, our \emph{Doppler Shift Test} can be effective without any specialized hardware, and it can be applied to existing devices.
\end{abstract}

\section{Introduction}\label{sec:intro}

As wireless communications enable an ever-broadening spectrum of mobile computing applications, \emph{location} or \emph{position} information
becomes increasingly important for those systems. Devices need to determine their own position,\footnote{In this paper, we are not concerned with the related but orthogonal localization problem of allowing a specific entity to determine and ascertain the location of other devices.} to enable location-based or location-aware functionality and services. Examples of such systems include: sensors reporting environmental measurements; cellular telephones or portable digital assistants (PDAs) and computers offering users information and services related to their surroundings; mobile embedded units, such as those for Vehicular Communication (VC) systems seeking to provide transportation safety and efficiency; or, merchandize (container) and fleet (truck) management systems.

\emph{Global navigation satellite systems} (GNSS), such as the Global Positioning System (GPS), its Russian counter-part (GLONAS), and the upcoming European GALILEO system, are the most widely used positioning technology. GNSS transmit signals bearing reference information from a constellation of satellites; computing platforms {\emph{nodes}), equipped with the appropriate receiver, can decode them and determine their own location.

However, commercial instantiations of GNSS systems, which are within the scope of this paper, are open to abuse: An adversary can influence the location information, $loc(V)$, a node $V$ calculates, and compromise the node operation. For example, in the case of a fleet management system, an adversary can
target a specific truck. First, the adversary can use a transmitter of forged GNSS signals that overwrite the legitimate GNSS signals to be received by the victim node (truck) $V$. This would cause a false $loc(V)$ to be calculated and then reported to the fleet center, essentially concealing the actual location of $V$ from the fleet management system. Once this is achieved, physical compromise of the truck (e.g., breaking into the cargo or hijacking the vehicle) is possible, as the fleet management system would have limited or no ability to protect its assets.

This is an important problem, given the consequences such attacks can have. In this paper, we are concerned with methods to mitigate such a vulnerability. In particular, we propose mechanisms to detect and reject forged GNSS messages, and thus avoid manipulation of GNSS-based positioning. Our investigation is complementary to cryptographic protection, which commercial GNSS systems do not currently provide but are expected to do so in the future (e.g., authentication services by the upcoming GALILEO system \cite{hein}). Our approach is motivated by the fundamental vulnerability of GNSS-based positioning to \emph{replay} attacks \cite{iwssc}, which can be mounted even against cryptographically protected GNSS.

The contribution of this paper consists of three mechanisms that allow receivers to detect forged GNSS messages and fake GNSS signals. Our countermeasures rely on information the receiver obtained before the onset of an attack, or more precisely, before the suspected onset of an attack. We investigate mechanisms that rely on own (i) \emph{location} information, calculated by GNSS navigation messages, (ii) \emph{clock} readings, without any re-synchronization with the help of the GNSS or any other system, and (iii) received GNSS signal \emph{Doppler shift} measurements. Based on those different types of information, our mechanisms can detect if the received GNSS signals and messages originate from adversarial devices. If so, location information induced by the attack can be rejected and manipulation of the location-aware functionality be avoided. We clarify that the reaction to the detection of an attack, and mechanisms that mitigate unavailability of legitimate GNSS signals is out of the scope of this paper.


We briefly introduce the GNSS operation and related work in Sec.~\ref{sec:gnss}. We discuss the adversary model and specific attack methods in Sec.~\ref{sec:attacks1}. We then present and analyze the three defensive mechanisms in Sec.~\ref{sec:defense}. Our findings support that highly accurate clocks can be very effective at the expense of appropriate clock hardware; but they can otherwise be susceptible, when off-the-shelf hardware is used. Location-based mechanisms can also be defeated relatively easily. On the contrary, our \emph{Doppler Shift Test} (DST) provides accurate detection of attacks, even against a sophisticated adversary.

\section{GNSS Overview}\label{sec:gnss}

\subsection{Basic Operation}

Each GNSS-equipped node $V$ can receive simultaneously a set of navigation messages $NAV_{i}$ from each satellite $S_i$ in the visible \emph{constellation}. Satellite transmitters utilize a spread-spectrum technique and each satellite is assigned a \emph{unique spreading code} $C_{i}$. These codes are \emph{a priori} publicly known. Navigation messages allow $V$ to determine its position, $loc(V) = (X_V,Y_V,Z_V)$, in a Cartesian system, as well global time, by obtaining a clock correction or \emph{time offset}, $t_{V}$, also called the \emph{synchronization error}. At least four satellites should be visible in order for a receiver to compute position and exact time, the so-called \emph{PVT} (Position, Velocity and Time) or \emph{navigation solution} \cite{kaplan}. This computation relies on the \emph{pseudo-range} measurements performed by $V$, one pseudo-range per visible satellite, that is, estimating the satellite-receiver distance based on the estimated signal propagation delay, $\rho_i$. For each pseudo-range $\rho_i$ estimated at $V$, the following equation is formed:
\begin{equation}
\rho_i=|s_i-loc(V)|+c \cdot t_V
\end{equation}
The satellite $S_i$ position is $s_i$, the receiver position is $loc(V)$, $c$ is the speed of light, and $t_V$ is the synchronization
error for $V$.

\subsection{Future Cryptographic GNSS Protection}

Cryptographic protection ensures the authenticity and integrity of GNSS messages, i.e., ensures that NAV messages generated solely by GNSS entities, with no modification, are accepted and used by nodes. Currently, cryptography is used in military systems, but it is not available for commercial systems to provide authenticity and integrity. \emph{Public or asymmetric key cryptography} is a flexible and scalable approach that does not require tamper-resistant receivers.\footnote{To prevent the compromise of a single, system-wide symmetric key, shared among the GNSS and all nodes.} Independently of the number of receivers present in the system (possibly, millions or eventually hundreds of millions), a pair of private/public keys $k_i,K_i$ can be assigned to each satellite $S_i$, with the public key bound to the satellite identity via a certificate provided by a Certification Authority. Each receiver obtains the certified public keys of all satellites in order to be able to validate NAV messages digitally signed with the corresponding $k_i$. \emph{Navigation Message Authentication} (NMA) \cite{hein} will be available as a GALILEO service.

To further enhance protection, a different public-key NMA approach was proposed in \cite{Kuhn}. Each $S_i$ chooses a secret spreading code for each NAV message but discloses this, along with a \emph{hidden timing marker}, in a delayed and authenticated manner to the receiving nodes. If nodes can maintain accurate clocks by means other than the GNSS system alone, they can then safely detect messages that are forged or replayed between the time of their creation and the code disclosure. A similar idea using \emph{Secret Spreading Codes} (SSC) was presented in \cite{scott}.

\section{Attacking GNSS}\label{sec:attacks}

\subsection{Adversary model}

The location (position) GNSS-equipped nodes obtain can be manipulated by an \emph{external} adversary, without any adversarial control on the GNSS entities (the system ground stations, the satellites, the ground-to-satellite communication, and the receiver). If any cryptographic protection is present, we assume that cryptographic primitives are not breakable and that the private keys of satellites cannot be compromised. The adversary can receive signals from all available satellites (depending on the locations of the adversary-controlled receivers). It is also fully aware of the GNSS implementation specifics and thus can produce fully compliant signals, i.e., with the same modulation, transmission frequency equal to the nominal one, $f_t$, or any frequency in the range of received ones, $f_r$; similarly, transmitted and received signal powers, as well as message preambles and body format (header, content).

We classify adversaries based on their ability to reproduce GNSS messages and signals, considering ones equipped with:
\begin{enumerate}
\item Single or multiple radios, each transmitting at the same constant power, $P^c_t$, and frequency $f^c_t$.
\item Single or multiple radios, each being ability to adapt its transmission frequency, $f^j_{t}$, over time; $j$ is an index of adversarial radios.
\item Multiple radios with adaptive transmission capabilities as above, and additionally the ability to establish fast communication among any of the adversarial nodes equipped with those radios.
\end{enumerate}

Adversarial radios in all above cases can \emph{record} GNSS signals and navigation messages for long periods. For all adversaries above, we consider a nominal range $R$, within which adversarial transmissions can be received, with this value varying for different adversarial radios. We denote this as the \emph{area under attack}. Clearly, the more powerful and the more numerous radios an adversary has, the higher its potential impact can be. In the sense, it can influence a larger system area and potentially mislead more receivers.

We assume that the area under attack does not coincide with the wireless system area. In other words, the adversary has limited physical presence and communication capabilities. This implies that nodes can lock on actual GNSS signals for a period of time before entering an area under attack. We do not dwell on how frequently and under what circumstances nodes are under attack. Rather, we investigate the strength of different defense mechanisms given that a node is under attack. We abstract the physical properties of the adversarial equipment and consider the periods of time it can cause unavailability and maintain the receiver locked on the spoofed signal.

We emphasize that our attack model is \emph{not} the worst case; this would be a receiver under attack during its \emph{cold start}, that is, the first time it is turned on and searches for GNSS signals to lock on. However, our adversary model corresponds to a broad range of realistic cases and it is a powerful one. For example, returning to the cargo example of the introduction: It will be hard for an adversary to control a receiver from its installation, e.g., on a container, and then throughout a trip. But it would be rather easy to select a location and time to mount its attack. Regarding the strength of the attacker, it is noteworthy that attacks are possible without any physical access to and without tampering with the victim node(s) software and hardware.


\subsection{Mounting Attacks against GNSS Receivers}\label{sec:attacks1}


The adversary can construct a transmitter that emits signals identical to those sent by a satellite, and mislead the receiver that signals originate from a visible satellite. However, the attacker has to first force the receiver to lose its ``lock'' on the satellite signals. This can be achieved by \emph{jamming} legitimate GNSS signals, by transmitting a sufficiently powerful signal that interferes with and obscures the GNSS signals \cite{volpe}. Jammers are simple to construct with low cost and very effective: for example, with 1 Watt of transmission power, the reception of GNSS signals is stopped within a radius of approximately 35 km radius \cite{kaplan,volpe}.

Then, the adversary can \emph{spoof} GNSS signals, i.e., forge and transmit signals at the same frequency and with power that \emph{exceeds} that of the legitimate GNSS signal at the receiver's antenna. Satellite simulators are capable of broadcasting simultaneously signals carrying counterfeit navigation data from ten satellites.\footnote{The adversary can deceive the receiver after down-conversion of the satellite signal, with one component in-phase and one in-quadrature:
\begin{equation}
I(t)=a_{i}C_{a}(t)M(t)cos(ft)
\end{equation}
\begin{equation}
Q(t)=a_{q}C_{a}(t)M(t)sin(ft)
\end{equation}
$C_{a}$ is the C/A (Course/Aquisition) code, $M(t)$ is the NAV message, and coefficients $a_{i}$ and $a_{q}$
represent the signal attenuation. The attacker could pick the amplifying coefficients $a_{i}$ and $a_{q}$ such that the received signal power
exceeds the nominal power od a GPS signal~\cite{spoofcon}.} The spoofed signal can also be generated by manipulating and rebroadcasting actual signals (\emph{meaconing}). As long as the lock of the victim receiver $V$ on the spoofed signal persists, $loc(V)$ is under the influence or full control of the adversary.

Apart from jamming, the adversary could take advantage of \emph{gaps in coverage}, i.e., areas and periods of time for which $V$ cannot lock on to more than three satellite signals. Clearly, this can be often possible in urban areas or because of the terrain, such as tunnels or obstructions from high-rise buildings. We do not consider further this case, as such loss of satellite signals is not under the control of the attacker. Nonetheless, the tests we propose here are effective independently of what causes receivers to loose lock on GNSS signals.


\subsection{Replay attack}



\begin{figure}[t]
\centering
\includegraphics[width=\linewidth]{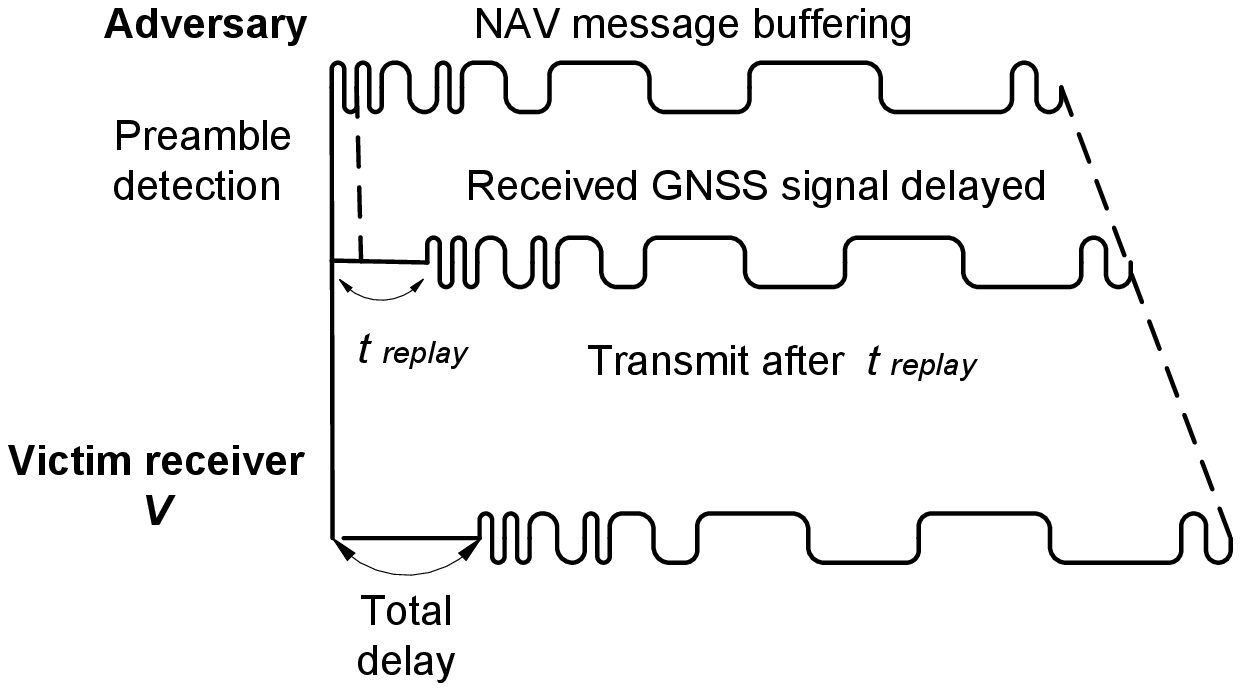}
\caption{Illustration of the replay attack: the adversary captures and
replays the signal after some time $t_{replay}=t^{min}_{replay}+\tau$,
with the $\tau \ge 0$ chosen by the adversary, and
$t^{min}_{replay}>0$ imposed by the specifics of the attack
configuration and the adversary capabilities.}\label{fig:delay}
\end{figure}

The \emph{replay attack} can be viewed as a part of a more general class of \emph{relay attacks}: the attacker receives at one location legitimate GNSS signals, relays those to another location where it retransmits them without any modification. This way the adversary can avoid detection if cryptography is employed, while it can ``present'' a victim with GNSS signals that are not normally visible at the victim's location. In this paper, we abstract away the placement of adversarial nodes, and we characterize the replay attack by two features: (i) the adversarial node capability to receive, record and replay GNSS signals, and (ii) the delay $t_{replay}$ between reception and re-transmission of a signal.

The GNSS signal reception and replay can be done at the message or symbol level, or it can be done by recording the entire frequency band and replaying it without de-spreading signals. The latter, more involved and thus costly, would enable the attacker to mount an attack against the delayed-disclosure secret spreading code approach, as pointed out in \cite{Kuhn}, not only for long replaying delays but also for very short ones. Clearly, such an instantiation of the replaying attack implies a more sophisticated adversary than one replaying symbols or messages. For example, the adversary would need to infer, possibly by possessing a legitimate receiver, the start of NAV messages to replay signals accordingly

The $t_{replay}$ delay between reception and re-transmission depends on the attack configuration (e.g., the distance between the receiving and re-transmitting adversarial radios, the physics of the signal propagation, and, when applicable, the delay for the adversary to decode the GNSS signal). We capture such factors by considering $t^{min}_{replay} > 0$, a minimum delay that the adversary cannot avoid. Beyond this, the attacker can choose some additional delay $\tau \ge 0$, such that it replays the signal after $t_{replay} = t^{min}_{replay} + \tau$. We illustrate a replay attack in Fig.~\ref{fig:delay}: The recording of the NAV message starts after its beginning is detected, due to the preamble 10001011, with length of eight chips, and the decoding of the NAV message first bit. This corresponds to $t^{min}_{replay}=20ms$: the transmission rate of 50 bit/s implies that 20ms are needed for the first bit to be received by an adversarial radio.


\begin{figure}[t]
\centering
\includegraphics[width=\linewidth]{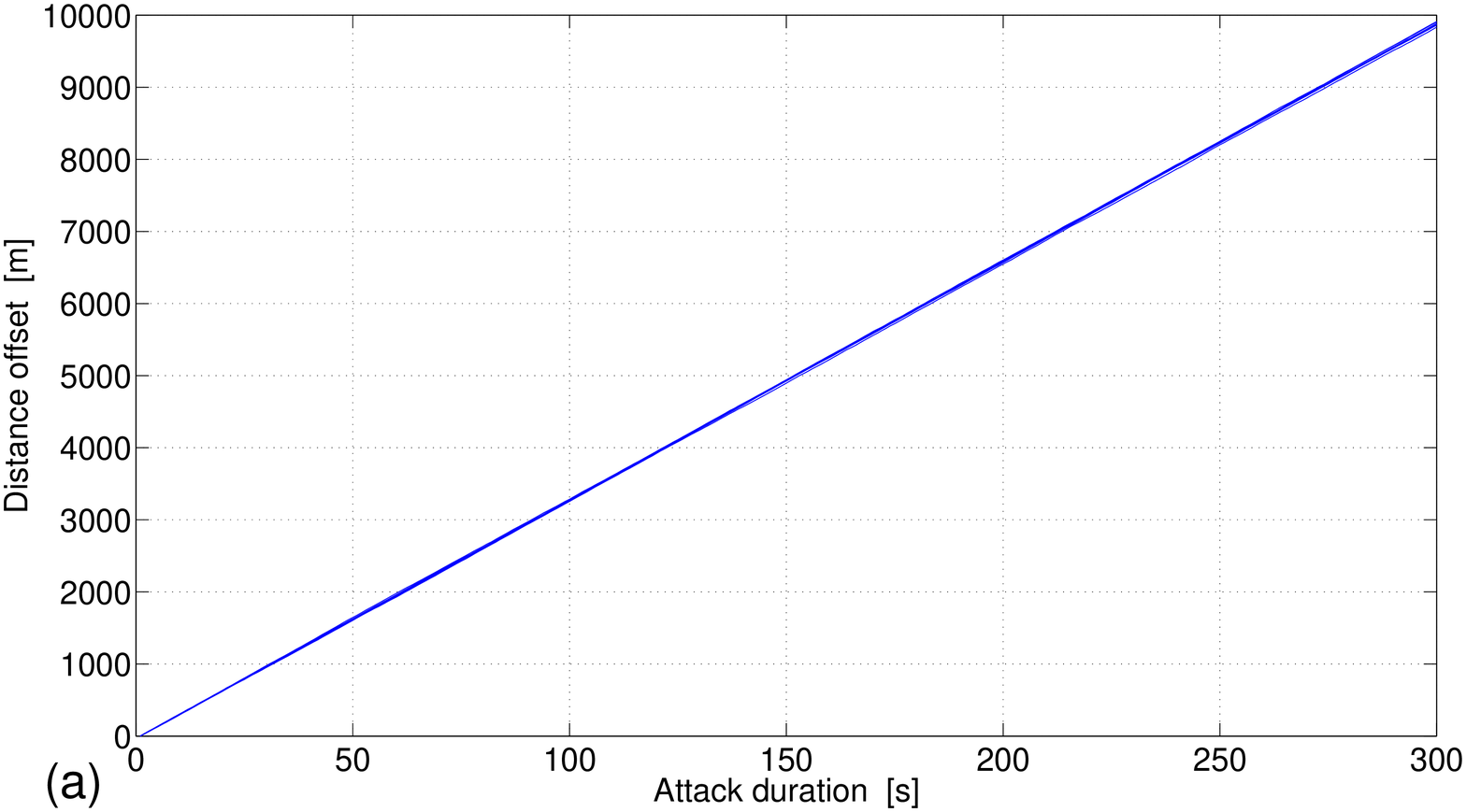}\hfill
\includegraphics[width=\linewidth]{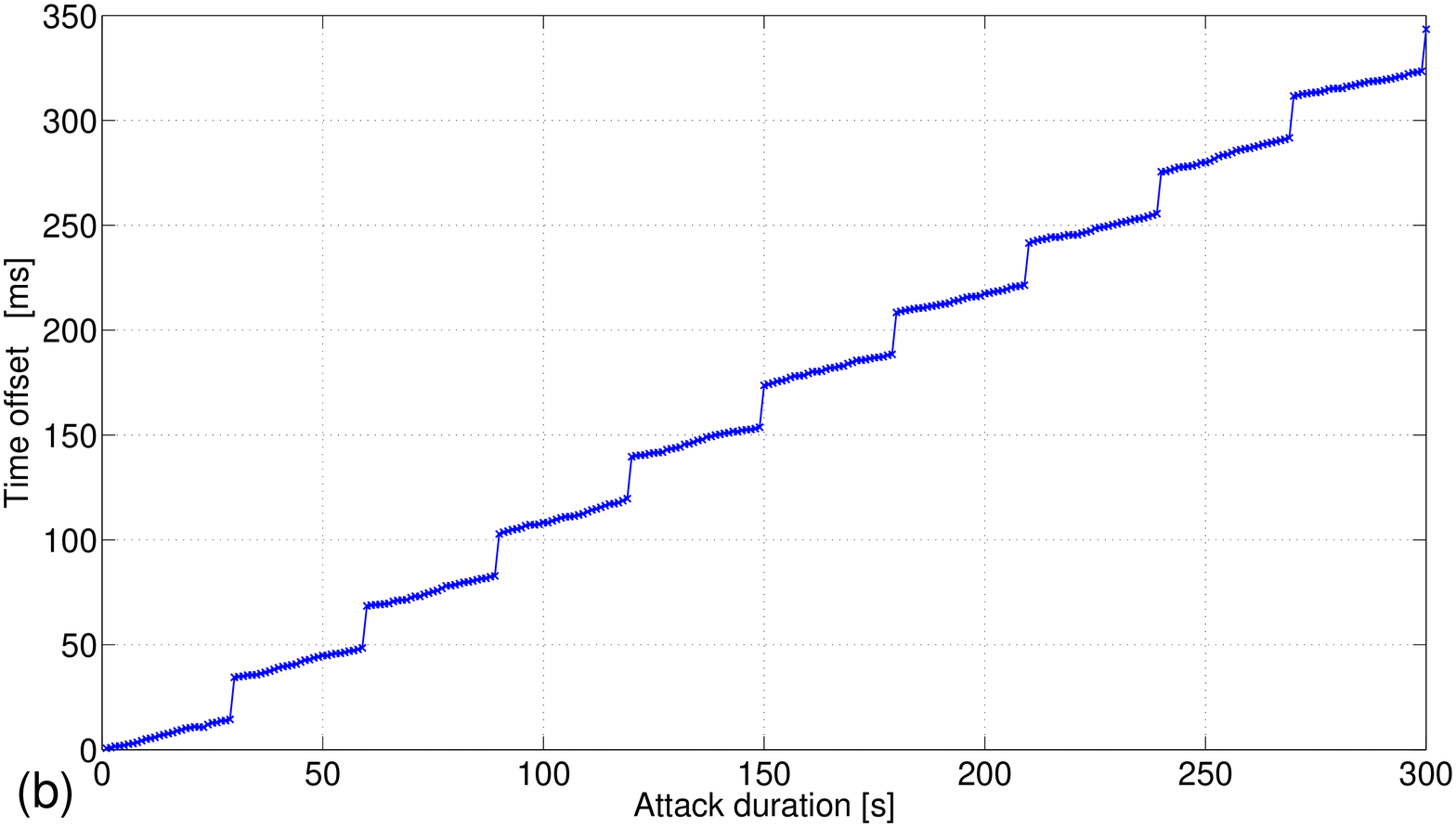}
\caption{Impact of the replay attack, as a function of the \emph{spoofing} attack duration. (a) Location offset or error: Distance between the attack-induced and the actual victim receiver position. (b) Time offset or error: Time difference between the attack-induced clock value and the actual time.}
\label{fig:reppos}
\end{figure}

The adversary can choose different $t_{replay}$ values for signals from different satellites, even though ``blind'' replaying of all NAV signals with the same delay can be effective. The selection of which signals (from which satellites) to relay offer flexibility. But even the ``blind'' replaying of all NAV signals (the entire band) can be effective: $t_{replay}$ controls the ``shift'' in the PVT solution. Essentially, $t_{replay}$ controls the ``shift'' in the PVT solution the adversary induces to the victim node(s).

Fig.~\ref{fig:reppos} shows the impact of a replay attack as a function of the \emph{spoofing} stage of the attack: (i) the \emph{location offset} or error, i.e., the distance between the attack-induced and the actual victim receiver position, and (ii) the \emph{time offset} or error, that is, the time difference between the attack-induced clock value and the actual time. We consider for this example $t_{relay}=20$ms, as the first bit decoding delay dwarfs the preamble detection and propagation delays. This is indeed a very subtle attack we refer to \cite{iwssc} for a range of $t_{replay}$ values, which shows that the larger the $t_{replay}$, as the adversary tunes its $\tau$ value, the higher the location and time offsets.

Even for a very low $t_{replay}$, while the mobile node receiver is still locked on the attacker-transmitted signals, the location error increases, with the victim receiver ``dragged'' away from its actual position. Each millisecond of $t_{relay}$ translates approximately into 300m of location offset for each pseudorange (as the speed of light, $c$, is taken into account), with the actual ``displacement'' of the victim depending on the geometry (e.g., position of the satellite whose signals were replayed).

As for the time offset, which can be viewed as a side-effect of the attack: it is in the order of less than one millisecond per second, and it can very well go easily unnoticed by the user. With a given $t_{relay}$, every time the victim receiver re-synchronizes, typically at the end of a NAV message that lasts $30$ sec, $t_{replay}$ will emerge as $t_V$ from the PVT solution and thus will be accumulated as part of the time offset shown in Fig.~\ref{fig:reppos}. 
\section{Defense mechanisms}\label{sec:defense}

We investigate three defense mechanisms that rely on a common underlying three-step idea. First, the receiver collects data for a given parameter during periods of time it deems it is not under attack; we term this the \emph{normal mode}. Second, based on the normal mode data, the receiver \emph{predicts} the value of the parameter in the future. When it suspects it is under attack, it enters what we term \emph{alert mode}}. In this mode, the receiver compares the predicted values with the ones it obtains from the GNSS functionality. If the GNSS-obtained values differ, beyond a protocol-selectable threshold, from the predicted ones, the receiver deems it is \emph{under attack}. In that case, all PVT solutions obtained in alert mode are discarded. Otherwise, the suspected PVT solutions are accepted and the receiver reverts to the normal mode.


In this work, we consider three parameters: \emph{location}, \emph{time}, and \emph{Doppler Shift}, and we present the corresponding detection mechanisms, \emph{Location Inertial Test}, \emph{Clock Offset Test}, and \emph{Doppler Shift Test}. We emphasize again that all three mechanisms rely on the availability of prior information collected in normal mode. But they are irrelevant if the receiver starts its operation without any such information (i.e., a \emph{cold start}).

To evaluate the proposed schemes, we use GPS traces collected by an ASHTECH Z-XII3T receiver that outputs observation and navigation (.obs and .nav) data into RINEX (\emph{Receiver Independent Exchange Format}) \cite{Office}. We implement the PVT solution functionality in Matlab, according to the receiver interface specification \cite{Office}. Our implementation operates on the RINEX data, which include pseudoranges and Doppler frequency shift and phase measurements. We simulate the movement of receivers over a period of $T=300s$, with their position updated at steps of $T_{step}= 1 sec$.

\subsection{Location Inertial Test}\label{sec:inertials}

At the transition to alert mode, the node utilizes own location information obtained from the PVT solution, to predict positions while in attack mode. If those positions match the suspected as fraudulent PVT ones, the receiver returns to normal mode. We consider two approaches for the location prediction: (i) inertial sensors and (ii) Kalman filtering.

\emph{Inertial sensors}, i.e., altimeters, speedometers, odometers, can calculate the node (receiver) location independently of the GNSS functionality.\footnote{They have already been used to provide continuous navigation between the update periods for GNSS receivers, which essentially are discrete-time position/time sensors with sampling interval of approximately one second} However, the accuracy of such (electro-mechanical) sensors degrades with time. One example is the low-cost inertial MEMS Crista IMU-15 sensor (Inertial Measurement Unit).

\begin{figure}[t]
\centering
\includegraphics[width=\linewidth]{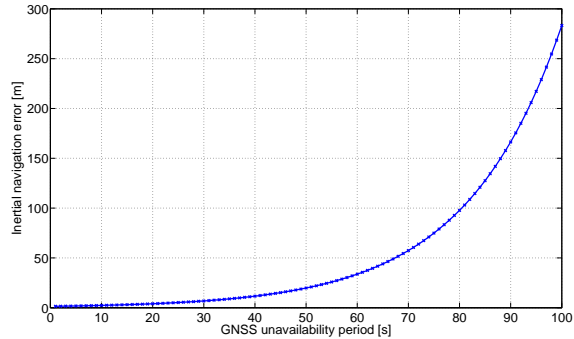}
\caption{Location error of Crista IMU-15 inertial sensor, as a function of the GNSS unavailability period.}\label{fig:pdferror}
\end{figure}

\begin{figure}[t]
\centering
\includegraphics[width=\linewidth]{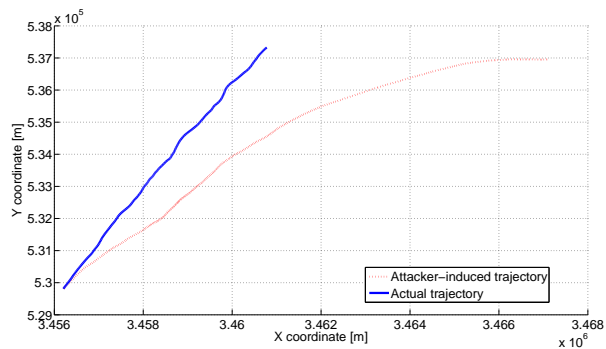}
\caption{Illustration of location error using inertial sensors: Actual vs. estimated when under attack trajectory.}\label{fig:posoffset}
\end{figure}

Fig.~\ref{fig:pdferror} shows the position error as a function of time \cite{calgary}, which is in our context corresponds to the period the receiver is in the alert mode. As the inertial sensor inaccuracy increases, the node has to accept as normal attack-induced locations. Fig.~\ref{fig:posoffset} shows a two-dimensional projection of two trajectories, the actual one and the estimated and erroneously accepted one. We see that over a short period of time, a significant difference is created because of the attack.


\begin{figure}[t]
\centering
\includegraphics[width=\linewidth]{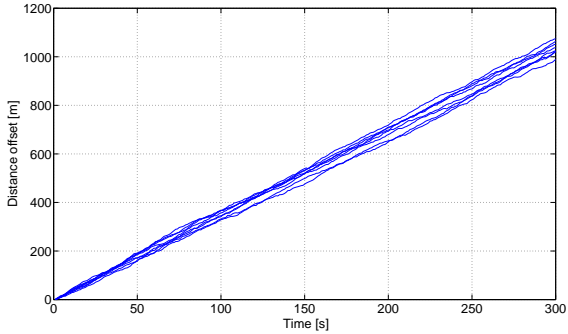}
\caption{Distance error of inertial mechanisms with Kalman filtering, as a function of the GNSS unavailability period.}\label{fig:new}
\end{figure}

A more effective approach is to rely on Kalman filtering of location information obtained during normal mode. Predicted locations can be obtained by the following system model:
\begin{equation}
S_{k+1}= \Phi_{k} S_{k}+W_{k}
\end{equation}
with $S_{k}$ being the system state, i.e., location $(X_{k},Y_{k},Z_{k})$ and velocity $(Vx_{k},Vy_{k},Vz_{k})$ vectors, $\Phi_{k}$ the transition matrix, and $W_{k}$ the noise. Fig.~\ref{fig:new} illustrates the location offset for a set of various trajectories. Unlike the case that only inertial sensors are used, with measurements of inertial sensors (with the error characteristics of Fig.~\ref{fig:pdferror} used as data when GNSS signals are unavailable, filtering provides a linearly increasing error with the period of GNSS unavailability.

Overall, for short unavailability periods, inertial mechanisms can be effective. As long as the error (Y axes of Figs.~\ref{fig:posoffset},~\ref{fig:new}) does not grow significantly, the replay attack can be detected. But for sufficiently high errors, the replay attack impact can remain undetected. We remind the reader that the x-axes in Fig.~\ref{fig:reppos} provide the duration of the spoofing attack - the transmission (replay) of GNSS signals - and they are not to be confused with the duration of the GNSS period of unavailability in the x-axis of  Figs.~\ref{fig:posoffset},~\ref{fig:new}.

\subsection{Clock Offset Test}


Each receiver has a clock that is in general imprecise, due to the drift errors of the quartz crystal. If the reception of GNSS signals is disrupted, the oscillator switches from normal to holdover mode. Then, the time accuracy depends only on the stability of the local oscillator~\cite{kaplan,franz}. The quartz crystals of different clocks run at slightly different frequencies, causing the clock values to gradually diverge from each other (skew error).



A simulation based study~\cite{franz} of quartz clocks claims that coarse time synchronization can be maintained at \emph{microsecond accuracy} without GPS reception for 350 sec in $95\%$ cases. This means that quartz oscillators can maintain millisecond synchronization for few hours, including random
errors and temperature change inaccuracies. Indeed, in such a case, the adversary would need to cause GNSS availability for long periods of time, for example, tens of hours, before being able to mount a relay attack that causes a time offset in the order of tens of milliseconds.

\begin{figure}[t]
\centering
\includegraphics[width=\linewidth]{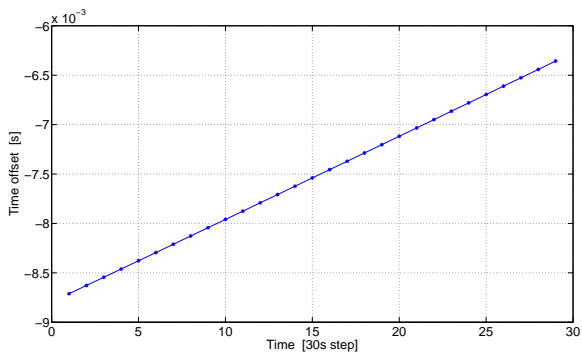}
\caption{Clock offset for the ASHTECH Z-XII3T receiver, during a 900 sec period with no re-synchronization.}\label{fig:offset}
\end{figure}

However, without highly stable clocks, mounting attacks against the Clock Offset Test can be significantly easier. This can be the
case for a ASHTECH receiver, for which time offset values are shown at successive points in time, each 30 seconds apart, in
Fig.~\ref{fig:offset}. We clarify this is not to be perceived as criticism for a given receiver or to be the basis for the suitability
of the Clock Offset Test. As explained above, the stability of the receiver clock determines the strength of this test. But the data in
Fig.~\ref{fig:offset}, over a period of 900 seconds, exactly demonstrates that for commodity receivers significant instability is observed; time offset values are in the order of ten milliseconds (or slightly less). Consequently, the adversary would need to jam for roughly a couple of minutes, force the receiver to consider as acceptable a time offset of 20 to 32 milliseconds, and thus be mislead by a replay attack as detailed in Sec.~\ref{sec:attacks}.

Finally, we note that we do not consider here the case of synchronization by means external to the GNSS system. For example, if the receiver could connect to the Internet and run NTP, it could obtain accurate time. But this would be an infrequent operation (in the order of magnitude of days), thus useful only if
highly stable clock hardware were available.

\subsection{Doppler Shift Test (DST)}

Based on the received GNSS signal Doppler shift, with respect to the nominal transmitter frequency ($f_{t}=1.575$GHz), the receiver can
predict future Doppler Shift values. Once lock to GNSS signals is obtained again, predicted Doppler shift values are compared to the ones calculated due to the received GNSS signal. If the latter are different than the predicted ones beyond a threshold, the GNSS signal is deemed adversarial and rejected. What makes this
approach attractive is the smooth changes of Doppler shift and the ability to predict it with low, essentially constant errors over long
periods of time. This in dire in contrast to the inertial test based on location, whose error grows exponentially with time.

The Doppler shift is produced due to the relative motion of the satellite with respect to the receiver. The satellite velocity is
computed using \emph{ephemeris} information and an orbital model available at the receiver. The received frequency, $f_{r}$, increases as
the satellite approaches and decreases as it recedes from the receiver; it can be approximated by the classical Doppler
equation:
\begin{equation}
f_{r}=f_{t} \cdot (1-\frac{v_{r} \cdot a}{c})
\end{equation}
where $f_{t}$ is nominal (transmitted) frequency, $f_{r}$ received frequency, $v_{r}$ is the satellite-to-user relative velocity vector
and $c$ speed of radio signal propagation. The product $v_{r} \cdot a$ represents the radial component of the relative velocity vector
along the line-of-sight to the satellite.

If the frequency shift differs from the predicted shift for each visible satellite $S_{i}$ in the area depending on the data obtained from the
almanac (in the case when the navigation history is available), for more than defined thresholds $(\Delta f_{min}, \Delta f_{max})$ or
estimated Doppler shift from navigation history differs for more than the estimated shift, knowing the rate ($r$), the receiver can deem
the received signal as product of attack.


The \emph{Almanac} contains approximate position of the satellites,
$(Xs_{i},Ys_{i},Zs_{i})$, time and the week number $(WN,t)$, and the
corrections, such that the receiver is aware of the expected
satellites, their position, and the Doppler offset.


Because of the high carrier frequencies and large satellite
velocities, large Doppler shifts are produced $(\pm 5 $kHz), and vary
rapidly (1 Hz/s). The oscillator of the receiver has frequency shift
of $\pm 3 $KHz, thus the resultant frequency shift goes therefore up
to $\pm 9$KHz. Without the knowledge of the shift, the receiver has
to perform a search in this range of frequencies in order to acquire
the signal. The rate of Doppler shift receiving frequency caused by
the relative movement between GPS satellite and vehicles
approximately 40 Hz per minute to the maximum. These variations are
linear for every satellite. If the receiver is mobile, the Doppler
shift variation can be estimated knowing the velocity of the
receiver(~\cite{patent1}).

In our simulations, Doppler shift is analyzed for each available
satellite (number of available satellites varies). To be consistent
with results shown for other mechanisms, we present results for DST
for the 300sec period. 

\begin{figure}[t]
\centering
\includegraphics[width=\linewidth]{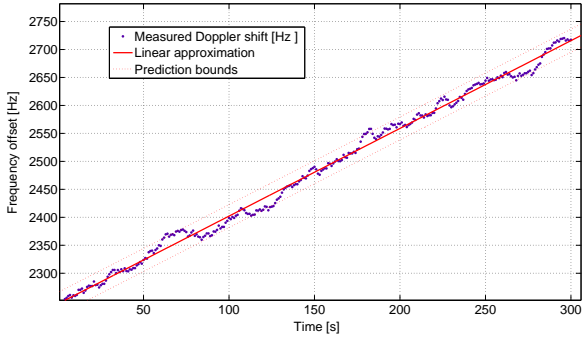}
\caption{Measured and approximated Doppler frequency shift.}\label{fig:bound1}
\end{figure}

We observe in Fig.~\ref{fig:bound1} the Doppler shift variation based on data collected by an ASHTECH receiver: the maximum change in rate is
within $+/- 20$Hz around a linear curve fitted to the data. This clues that with sufficient samples, the future Doppler Shift rate,
and thus the shift per se, values can be predicted. In practice, we observe that $50$ sec of samples, with one sample per second, appear to be sufficient.

More precisely, the rate of change of the frequency shift, $D_i(t)$, is computed for each satellite, $S_i$, as:
\begin{equation}
r_{i}=\frac{dD_i(t)}{dt}
\end{equation}
which can be approximated by numerical methods. Based on prior samples for each $D_i$, available for some time window the frequency shift
can be predicted based those samples and the estimate rate of change of the Doppler shift. Based on prior measured statistics of the signal at the receiver, the variance $\sigma^{2}$ of a random component, assumed to be $N(0,\sigma^{2})$, can be estimated. This random component is due to signal variation (including receiver mobility, RF multipath, scattering). Its estimation can serve to determine an acceptable interval around the predicted values.



\begin{figure}[t]
\centering
\includegraphics[width=\linewidth]{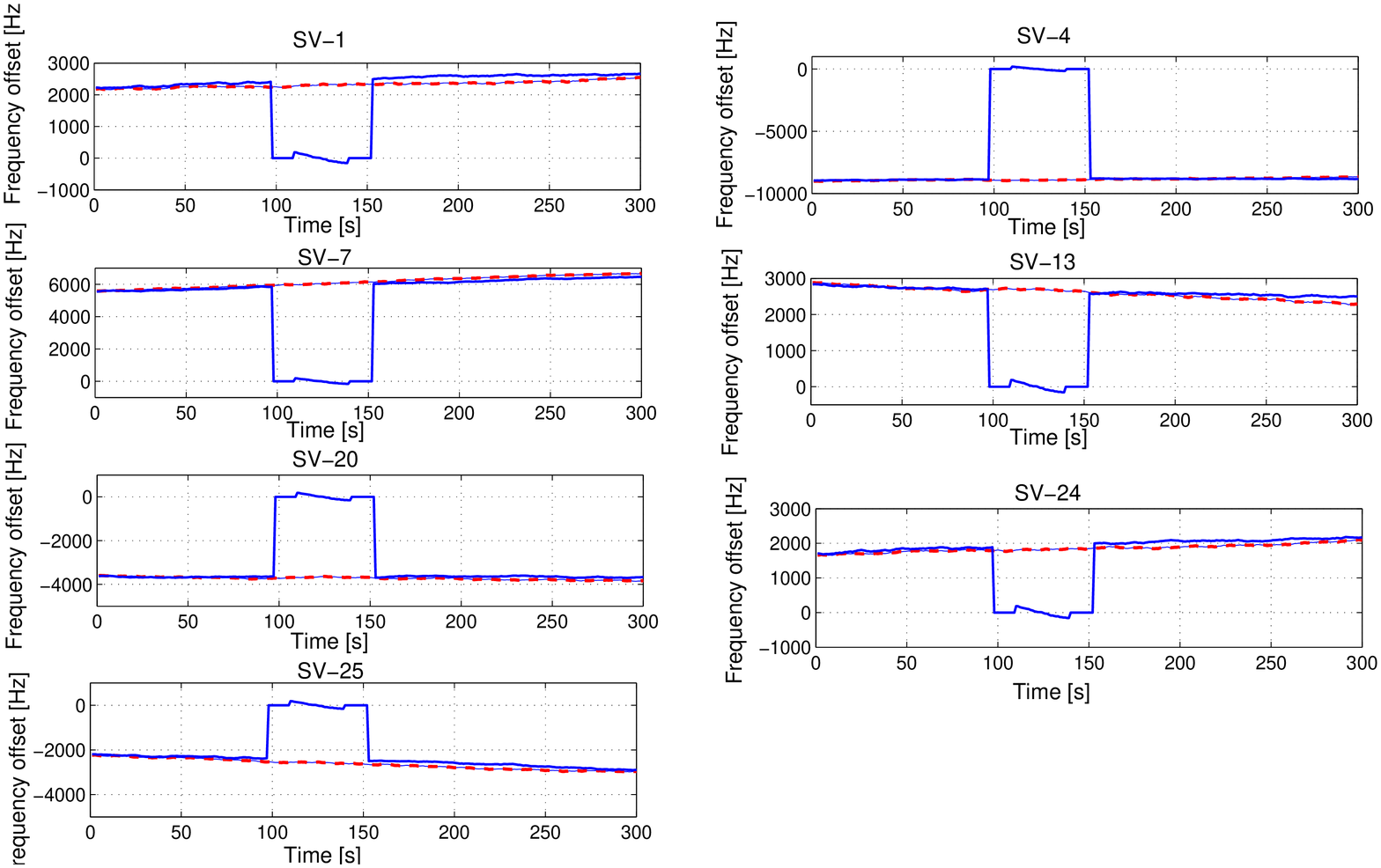}
\caption{Doppler shift attack; unsophisticated adversary. The dotted line represents the predicted and the solid line the measured frequency offset.}\label{fig:1}
\end{figure}

The adversary is mostly at the ground and static or moving with speed that is much smaller than the satellite velocity, which is in a range around 3km/s. Thus, the adversary will not be able to produce the same Doppler shift as the satellites, unless it changes its transmission frequency to match the one receivers would obtain from GNSS signals due to the Doppler shift. An unsophisticated attacker would then be easily detected. This is illustrated in Fig.~\ref{fig:1}: After a ``gap'' corresponding to jamming, there is a striking difference, between 100 and 150 seconds, when comparing the Doppler shift due to the attack to the predicted one.

\begin{figure}[t]
\centering
\includegraphics[width=\linewidth]{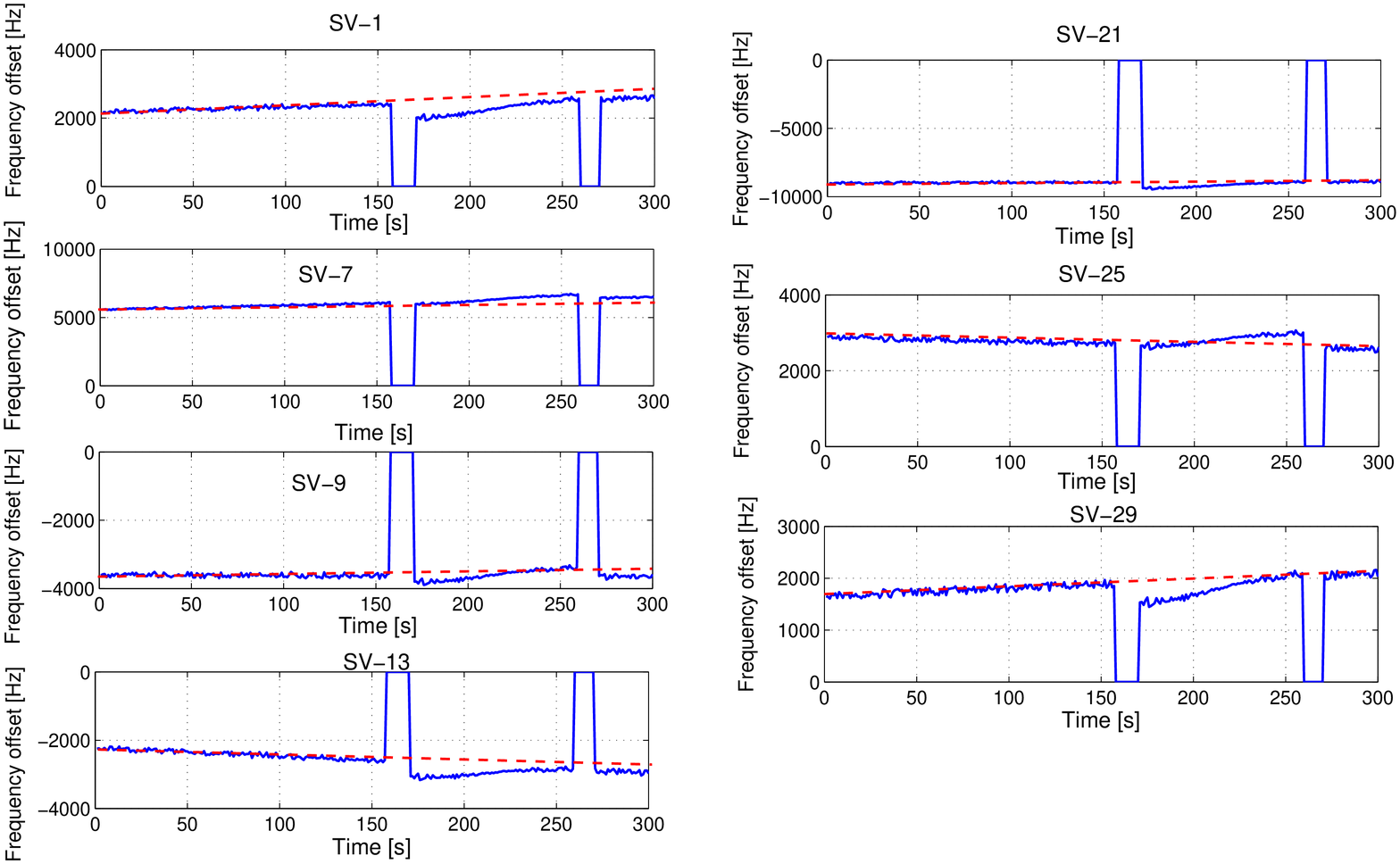}
\caption{Doppler shift attack; sophisticated adversary. The dotted line represents the predicted and the solid line the measured frequency offset.}\label{fig:11}
\end{figure}

The case of A sophisticated adversary that controls its transmission frequency (the attack starts at $160s$)is shown in the Fig.~\ref{fig:11}. The adversary has multiple adaptive radios and it operates according to the following principle: it predicts the Doppler frequency shift at the location of the receiver, and it then changes its transmission frequency accordingly. If the attacker is not precisely aware of the actual location and motion dynamics of the victim node (receiver), there is still a significant difference between the predicted and the adversary-caused Doppler shift. This is shown, with a magnitude of approximately 300 Hz, in Fig.~\ref{fig:11}; a difference that allows detection of the attack.


\section{Conclusion}

Existing GNSS receivers are vulnerable to a number of attacks that
manipulate the location and time the receivers compute. We
qualitatively and quantitatively analyze those in this paper, and
identify memory-based mechanisms that can help in securing GNNS
signals. In particular, we realize that location-based inertial
mechanisms and a clock offset test can be relatively easily defeated,
with the adversary causing (through jamming) a sufficiently long
period of unavailability. In the latter case, only specialized highly
stable clock hardware could enable detection of fraudulent GNSS
signals. Our Doppler Shift Test provides resilience to long
unavailability periods without specialized equipment.

Our results are the first, to the best of our knowledge, to provide
tangible demonstration of effective mechanisms to secure mobile
systems from location information manipulation via attacks against
the GNSS systems.

As part of on-going and future work, we intent to further refine and
generalize the simulation framework we utilized here, to consider
precisely the effect of counter-measures that only partially limit
the attack impact. Moreover, we will consider more closely the cost
of mounting attacks of differing sophistication levels, especially
through proof-of-concept implementations.

\end{document}